# Pulmonary surfactant inhibition of nanoparticle uptake by alveolar epithelial cells

M. Radiom[1,2*], M. Sarkis[1], O. Brookes[3], E.K. Oikonomou[1], A. Baeza-Squiban[3] and J.-F. Berret[1*]

[1]Université de Paris, UMR CNRS 7057, Laboratoire Matière et Systèmes Complexes, Paris, France.
[2]Present address: Institute for Food, Nutrition and Health, D-HEST, ETH Zürich, Zürich, Switzerland
[3]Université de Paris Unité BFA, UMR CNRS 8251, Laboratoire de Réponses Moléculaires et Cellulaires aux Xénobiotiques, Paris, France

Correspondence: milad.radiom@hest.ethz.ch, jean-francois.berret@u-paris.fr

**Abstract**
Pulmonary surfactant forms a sub-micrometer thick fluid layer that covers the surface of alveolar lumen and inhaled nanoparticles therefore come in to contact with surfactant prior to any interaction with epithelial cells. We investigate the role of the surfactant as a protective physical barrier by modeling the interactions using silica-Curosurf®-alveolar epithelial cell system *in vitro*. Electron microscopy displays that the vesicles are preserved in the presence of nanoparticles while nanoparticle-lipid interaction leads to the formation of mixed aggregates. Fluorescence microscopy reveals that the surfactant decreases the uptake of nanoparticles by up to two orders of magnitude in two models of alveolar epithelial cells, A549 and NCI-H441, irrespective of immersed culture on glass or air-liquid interface culture on transwell. Confocal microscopy corroborates the results by showing nanoparticle-lipid colocalization interacting with the cells. Our work thus supports the idea that pulmonary surfactant plays a protective role against inhaled nanoparticles. The effect of surfactant should therefore be considered in predictive assessment of nanoparticle toxicity or drug nanocarrier uptake. Models based on the one presented in this work may be used for preclinical tests with engineered nanoparticles.

## 1. Introduction

The exchange of gas in the lungs during breathing exposes the epithelium of the respiratory tract to pathogens and airborne particulate matter (PM) such as natural and engineered particles. Exposure to natural and anthropogenic PM such as crystalline silica dust, asbestos fibers, coal dust, and diesel exhaust particles has long been known to pose important health hazards, *e.g.* increasing the risk of lung cancer and chronic obstructive pulmonary disease (COPD). Therefore, inhalation of PM is a current global health concern [1]. Geographical maps of air pollution suggest a world average concentration of 25 µg m$^{-3}$ of suspended PM$_{2.5}$ (PM with an aerodynamic diameter ≤ 2.5 µm) [2,3]. This concentration is equivalent to 300 µg of PM$_{2.5}$ inhaled on daily basis, of which 8 – 50% or 20 – 160 µg may enter the lungs as schematically shown in Fig. 1a [4]. Movement of PM$_{2.5}$ to different regions of the lungs is dependent on several factors including physiological ones such as breathing patterns, and physical ones such as particle size and deposition mechanics [4]. Several models for estimating deposition fraction in different regions of the lungs exist, and they show that nanoparticles (aerodynamic diameter ≤ 100 nm) can spread deeply in the lungs [4]. The deposition fraction of nanoparticles in distal lung regions, *i.e.* the alveolar region, is estimated to be 15 – 32% from IDEAL model (for oral breathing, tidal volume = 1000 ml, and frequency of breathing = 15 min$^{-1}$) [4]. This model's estimations are reproduced in Fig. 1c. Using the above range and assuming exposure to nanoparticles alone, the lung burden of inhaled nanoparticles in the alveolar region is 3 – 50 µg on daily basis.

The alveolar region is composed of sponge-like clusters of pulmonary alveoli (air sacs), each of which has a diameter of about 200 µm, as shown schematically in Figs. 1a and 1b. The alveolar region presents a very large potential exposure surface, more than 70 m$^2$ [5]. Furthermore, the alveoli are bounded by an epithelial cell layer, which rests in contact with endothelial cells of pulmonary capillaries *via* an extracellular matrix. The tissue is very thin, less than 2 µm, and is a major potential route of systemic entry of inhaled nanoparticles [6-13]. For example, about 20 – 30% of endotracheally ventilated aerosolized titanium oxide nanoparticles (22 nm) had translocated through the alveolar lumen within 1 h post-exposure [8]. For intratracheally instilled gold nanoparticles (bare particle diameter equal to 11 nm, and with polyethylene glycol coating equal to 21 nm and 31 nm), up to 3% had entered secondary organs including liver, spleen and kidneys [14].

The first barrier to nanoparticles in the alveoli is pulmonary surfactant. This fluid is secreted by type II epithelial cells and covers the epithelium with a thickness of about 0.2 – 0.5 µm (Fig. 1b) [15]. The surfactant is composed of 92% lipids, out of which 78% are zwitterionic phosphatidylcholines (PC), and predominantly dipalmitoyl PC. The non-lipid fraction of surfactant primarily consists of proteins. Upon secretion, the surfactant forms a lipid monolayer at air-liquid interface, and tubular myelin and lamellar body structures in the film. Formation and stability of these structures is mediated by interactions with surfactant proteins, notably SP-A, SP-B and SP-C [16-18]. Surfactant proteins SP-A and SP-D are furthermore components of innate immune system and facilitate complement activation and phagocytosis [18]. The fate of nanoparticles in alveolar region, including translocation through the epithelial cell layer or internalization by immune cells [6,19-23], depends on their interaction with the surfactant.

These observations have prompted investigations into nanoparticle-surfactant interaction and its role on cellular uptake of the nanoparticles. Reported interactions with silica, alumina and latex nanoparticles in dilute surfactant include formation of supported lipid bilayers (SLBs), internalization by lipid vesicles, and aggregate formation. These assemblies are formed spontaneously and depend on particle size, roughness, and short- and long-range interactions such as steric and electrostatic forces [24-27]. Other non-specific interactions leading to disintegration, reduction in size and deformation of multivesicular structures have been observed with titanium oxide nanoparticles at *in vivo* surfactant concentration [28]. These interactions can change the biophysical properties of surfactant as recently shown in case of its flow properties [29].

Subsequent investigations examined cell responses to nanoparticles mediated by interactions with surfactant. Among other studies, it was shown by us that surfactant diminishes the uptake of silica nanoparticles by murine macrophages and lung carcinoma cells [30]. It was also found that surfactant reduces the amount of internalized dextran nanogels used for delivery of small interfering RNA [31,32]. The uptake of silver nanowires was however mediated only by harvested *in vitro* surfactant secretions which includes proteins, while no reduction was observed in the presence of Curosurf® (a surfactant substitute which lacks SP-A and SP-D) [33]. These differences of behavior are related with different physical and chemical properties of nanoparticles, *e.g.* metal *vs* non-metal, spherical *vs* ellipsoidal, and surfactant, *e.g.* containing proteins SP-A and SP-D, or devoid of them, in addition to cells with different uptake mechanisms, *e.g.* murine macrophages, human primary cells and cell lines, but also on relative concentration of surfactant to nanoparticles. Thereby our understanding of the fate of nanoparticles in pulmonary alveoli is not complete.

To this end, we mimic this process *in vitro* by exposing a monolayer of alveolar epithelial cells to a mixture of clinical surfactant substitute and nanoparticles. Our model nanoparticles are positively charged silica. Silica nanoparticles are commonly used in cosmetology products, in food as emulsifiers, in paints and as medical tools *e.g.* as drug-carriers, contrast agents and anti-cancer therapeutics [34,35]. The high level of consumption of this material has resulted in its mass production leading to a high risk of exposure by inhalation especially in occupational settings. (A list of production volume and industrial applications for silica nanoparticles is provided in **Supplementary Information S1**.) Long term exposure to crystalline silica dust is associated with pulmonary diseases such as silicosis, lung cancer, COPD and pulmonary tuberculosis [34]. In addition to silica dust, amorphous silica nanoparticles may exhibit toxicity depending on surface properties [36]; however, they pose an additional hazard associated with translocation through the alveolar tissue and systemic circulation leading to accumulation in secondary organs [6-14]. The choice of a positive surface charge is motivated by recent observations that nanoparticles with positive surface charge are more cytotoxic to macrophages as compared to neutral and negatively charged particles [37]. Furthermore, our group has recently shown that nanoparticles with positive surface charge interact strongly with pulmonary surfactant *via* electrostatic forces [26,38]. In a previous work, we focused our attention on SLBs that were induced on silica nanoparticles *via* sonication [39]. The current study investigates the protective role of surfactant in an attempt to mimic the *in vivo* exposure condition more closely. Thereby, silica nanoparticles and model surfactant Curosurf® were mixed without sonication. In this case, we did not obsreve SLBs around nanoparticles; however, nanoparticles were found inside the mutivesicular structures and adhering to the lipid membranes. The choice of Curosurf® is motivated by its approximation of the native surfactant in terms of lipid and protein compositions (**S2**). Curosurf® is also used in the treatment of respiratory distress syndrome (RDS) in pre-term enfants. We examined two cell lines, namely A549 and NCI-H441; both are common models of pulmonay alveolar cells while the latter shows better barrier forming properties. The investigations were performed on glass and on transwell. The choice of transwell is to induce air-liquid interface (ALI) which leads to tight epithelium forming in NCI-H441 cells. We measured uptake at several surfactant to nanoparticle concentration ratios, where we found reduction by two orders of magnitude in uptake when surfactant was present.

By fluorescently tagging the lipids [40] and the nanoparticles, we showed their colocalization on exposure to cells. Our result thus corroborates the protective role of surfactant in reducing the number of nanoparticle-cell interactions leading possibly to a reduced nanoparticle uptake, which further manifests the protective role of surfactant in reducing systemic circulation from alveolar entry route.

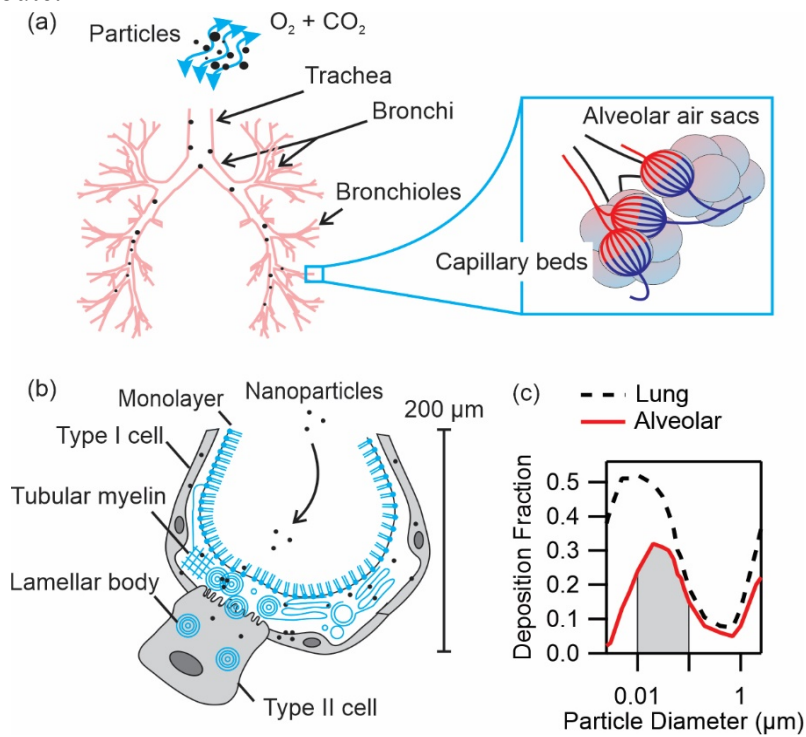

*Figure 1: (a) Schematic of particles entering the lungs during respiration. Lung sections including trachea, bronchi, bronchioles and alveolar air sacs are depicted. Airway diameters decrease in the same order from 12 mm to 200 µm. Nanoparticles can diffuse deeply and enter the alveolar air sacs. (b) Alveolar epithelium is composed of type I and type II cells. Type II cells secrete surfactant to alveolar lumen which then forms surfactant monolayer at air-liquid interface, tubular myelin and lamellar body structures in the surfactant film. (c) Total lung, i.e. tracheobronchial plus alveolar, and alveolar deposition fractions reproduced from IDEAL model calculations in Ref. [4]. Shaded area marks the deposition fraction of nanoparticles, diameter in the range 10 – 100 nm, in the alveolar region equal to 15 – 32%.*

2. **Material and Methods**

*Cell culture.* Adenocarcinoma A549 and NCI-H441 were obtained from the American Type Culture Collection (ATCC, USA). NCI-H441 was cultured in RPMI-1640 supplemented with 10% fetal bovine serum (FBS), 1% penicillin-streptomycin (PS), 1% GlutaMAX, 1% non-essential amino acids, 1% sodium pyruvate and 200 nM dexamethasone in T25 flasks inside an incubator (atmosphere of 5% $CO_2$, 37°C). Dexamethasone is a glucocorticoid supplement used to promote maturation of epithelial cells and to improve their barrier properties [41]. A549 was cultured in DMEM supplemented with 10% FBS and 1% PS in T25 flasks inside an incubator (5% $CO_2$, 37°C). Of note, A549 was not cultured in dexamethasone, except for the experiments on transwell and after establishment of air-liquid interface (ALI) culture, because it has been shown that dexamethasone inhibits A549 proliferation [42]. Both cell lines were harvested by trypsinization with Trypsin-EDTA. Sterilized glass coverslips were placed inside 6-well plates. Cells were seeded at a density of $4\times10^4$ cells cm$^{-2}$ and incubated for 48 h. Costar® 12-mm Transwell® 0.4-µm pore polyester membrane inserts were used for ALI culture. On transwell, cells were seeded at a density of $8\times10^4$ cells cm$^{-2}$. The cells were first incubated in the presence of their respective culture media in the apical and basal compartments of transwell inserts until they reached confluency. Thereafter, the culture state was changed to ALI. As of the beginning of ALI, dexamethasone was added to the basal medium of A549 cultures. RPMI-1640, DMEM, FBS, PS, GlutaMAX, and Trypsin-EDTA were all Gibco brand.

*Silica nanoparticles.* Positively charged silica nanoparticles were synthesized by Stöber method and used throughout this study [43]. The nanoparticles contain a core of Cy3 fluorescent molecules ($\lambda_{ex}$ = 550 nm, $\lambda_{em}$ = 570 nm) incorporated during fabrication. The nanoparticles have surface-grafted amine groups affording them a positive surface charge. The diamater of the nanoparticles as obtained from TEM is 41.2 nm and from DLS is 60 nm (pH < 6.0).

*Pulmonary surfactant.* Curosurf® (Chiesi Pharmaceuticals) was used as the surfactant model. This porcine-originated extract is generally used in replacement therapy to cure respiratory distress syndrome in preterm infants. The stock solution has a concentration of 80 mg ml$^{-1}$. Curosurf® lipids were labeled with the lipophilic fluorescent dye PKH67 ($\lambda_{ex}$ = 490 nm, $\lambda_{em}$ = 502 nm) at $10^{-3}$ µmol of PKH67 to 0.8 mg of Curosurf® following a previous protocol [40]. Curosurf® was provided by Dr. Mostefa Mokhtari (neonatology department of Kremlin-Bicêtre Hospital).

*Mixture preparation.* To prepare nanoparticle-Curosurf® dispersions, we define a mixing ratio parameter $X$ representing the relative concentration of Curosurf® to nanoparticles, namely $X = C_{\text{Curo}}/C_{\text{NP}}$. $X = 2$ and 20 were used for cell exposure experiments, while $X = 80$ was used for TEM investigations as explained in the following. Nanoparticle-Curosurf® mixture at $X = 2$ was prepared by first diluting the stock nanoparticle solution to $C_{\text{NP}} = 4$ mg ml$^{-1}$ in Milli-Q water and the stock Curosurf® solution to $C_{\text{Curo}} = 4$ mg ml$^{-1}$ in milli-Q water. Then, an appropriate volume of the diluted nanoparticle solution (50 µl for exposure tests on glass and 12.5 µl for exposure tests on transwell) was mixed with diluted Curosurf® solution at twice that volume. Thereafter, the total volume of the mixture was increased in serum free culture medium in order to obtain a nanoparticle concentration of $C_{\text{NP}} = 200$ µg ml$^{-1}$. During exposure to cells, the concentration was further reduced to $C_{\text{NP}} = 100$ µg ml$^{-1}$ (see *Exposure treatment*). Nanoparticle-Curosurf® mixture at a mixing ratio $X = 20$ was prepared similarly however the starting concentrations were 21 mg ml$^{-1}$. Then, an appropriate volume of the diluted nanoparticle solution (9.5 µl for exposure tests on glass and 2.4 µl for exposure tests on transwell) was mixed with the diluted Curosurf® solution at twenty times that volume. Thereafter, the total volume of the mixture was increased in serum free culture medium in order to obtain a nanoparticle concentration of 200 µg ml$^{-1}$. During exposure to cells, the concentration was further reduced to 100 µg ml$^{-1}$ (see *Exposure treatment*).

All sample preparations were performed immediately prior to cell exposure. The serum-free culture media were respectively RPMI-1640 and DMEM for NCI-H441 and A549. The total mixture volumes were respectively 2 ml and 500 µl in exposure tests on coverslip and on transwell. A summary of the exposure conditions is provided in Table 1, and the extended calculations in **S3** & **S4**. Finally, nanoparticle-Curosurf® mixture at $X = 80$ was prepared for TEM investigations. Here, 100 µl of Curosurf® stock solution (80 mg ml$^{-1}$) was mixed with 100 µl of nanoparticle solution at a concentration of 1 mg ml$^{-1}$.

**Table 1: Exposure concentrations on coverslip and transwell.**

| Exposure | $m_{NP}$ (µg) | $m_{Curo}$ (µg) | $V_{TOT}$ (mL) | $c_{NP}$ (µg ml$^{-1}$) | $m_{NP}/A$ (µg cm$^{-2}$) |
|---|---|---|---|---|---|
| Coverslip ($A = 9.6$ cm$^2$) | | | | | |
| Neat nanoparticle | 200 | 0 | 2.0 | 100 | 20.8 |
| X = 2 | 200 | 400 | 2.0 | 100 | 20.8 |
| X = 20 | 199.5 | 4000.5 | 2.0 | 100 | 20.8 |
| Neat Curosurf® | 0 | 400 | 2.0 | 0 | 0 |
| Transwell ($A = 1.12$ cm$^2$) | | | | | |
| Neat nanoparticle | 50 | 0 | 0.5 | 100 | 44.6 |
| X = 2 | 50 | 100 | 0.5 | 100 | 44.6 |
| X = 20 | 50.4 | 1000 | 0.5 | 100 | 44.6 |
| Neat Curosurf® | 0 | 100 | 0.5 | 0 | 44.6 |

*Exposure treatment.* Cultures were first rinsed with PBS to remove dead cells. Then, serum free medium was added to the cells to a volume of 1 ml for exposure tests on coverslip and 250 µl on transwell. Thereafter, an equal volume of mixture (see *Mixture preparation*) was added. The cells were incubated during 24 h in an atmosphere of 5% CO$_2$ at 37°C. Thereafter, the supernatant was discarded, and the cells were rinsed with PBS to remove remaining nanoparticles and Curosurf® vesicles, as well as the dead cells. Each exposure condition, namely neat nanoparticles, $X = 2$ and $X = 20$, was tested in duplicate.

*Cell fixation and immunostaining of nuclei.* The cells were fixed with 4% paraformaldehyde (PFA) in PBS, and then the nuclei were stained with DAPI. The cell-side of the coverslip was then wetted with PBS, and mounted and sealed to a glass slide using Gene Frame adhesive (ABgene Advanced Biotech). The transwell membranes were first cut using a scalpel, then the cell-side was wetted with Fluoroshield (Sigma-Aldrich), and mounted and sealed to a glass slide using a Gene Frame and a coverslip support.

*Barrier integrity immunolabelling and microscopy.* A549 and NCI-H441 were grown on transwell for 2 weeks, washed with PBS and fixed with 4% PFA in PBS for 10 minutes at ambient temperature. Inserts were washed, permeabilized with 0.1% Triton X-100 in PBS for 2 minutes, and then blocked with 2% FBS in PBS for 60 minutes at ambient temperature. Membranes were carefully excised and transferred to a humidified container for primary antibody binding. This was carried out overnight at 4°C, using a mixture of mouse anti-human ZO-1 and goat anti-human claudin 4 antibodies, both at 1:200 dilution in blocking buffer. Membranes were gently washed to remove unbound primary antibody, and were then incubated with a mixture of donkey anti-goat 594 and donkey anti-mouse 647 secondary antibodies for 60 minutes at ambient temperature. Samples were finally washed and mounted using soft mountant containing DAPI. Imaging was carried out on a Zeiss LSM 710, with the pinhole set to give 1 μm slices at each wavelength. Images shown are maximum projections of these stacks.

*Transepithelial electrical resistance.* The barrier integrity of the epithelia formed by NCI-H441 and A549 on transwell inserts was monitored by collecting the transepithelial electrical resistance (TEER). TEER was measured using EVOM2 Epithelial Volt/ohm meter along with Endohm12 chamber (World Precision Instrument) every 2 – 3 days as of day-0 of ALI. After sterilization with ethanol 70%, the chamber was rinsed and then filled with 2 mL of HBSS containing $Ca^{2+}$, $Mg^{2+}$ ions and glucose. The top compartment of transwell inserts was filled with 500 μL of HBSS and successively transferred in the chamber. The resistance value in $\Omega$ units was read after the stabilization of TEER signal. The value was subtracted by a background resistance due to transwell insert. The resulting value was subsequently multiplied by the surface area of transwell insert equal to 1.12 $cm^2$.

*Transmission electron microscopy.* A dispersion was prepared by simple mixing equal volumes of 80 mg $ml^{-1}$ Curosurf® and 1 mg $ml^{-1}$ silica nanoparticles. The dispersion was fixed with 2.5% glutaraldehyde and 4% PFA in PBS buffer. The sample was stored in the fixative for 3 h at room temperature and then kept at 4°C before next preparations. Afterwards, the sample was centrifuged at 500 – 1000 g for 3 min to obtain a pellet. After several washing steps in buffer, the sample was subsequently post-fixed in osmium tetroxide, dehydrated in an ascending ethanol series and embedded in Epon at 60°C. The Eppendorf cups were then removed, and ultrathin 70-nm sections were cut using an ultramicrotome UC6 (Leica). The sections were analyzed with a 120-kV transmission electron microscope Tecnai 12 (ThermoFisher Scientific) using the 4-k camera OneView and the GMS3 software (Gatan).

*Fluorescence microscopy.* Fluorescence microscopy images were obtained using an IX73 inverted microscope (Olympus) with a 10× objective (N.A. 0.3), an ExiBlue camera (QImaging), an XCite illumination system and MetaVue imaging software. The sample was illuminated, and the red fluorescence signal obtained *via* an excitation filter and an emission filter. The 2D signal intensity was averaged along the y-direction in the field of view, resulting in a 1D signal along the x-direction, using the Plot Profile plugin of ImageJ. For high intensity signals, *e.g.* in the case of exposure to neat nanoparticles, a background correction was applied: from the 1D signal, an intensity corresponding to cells alone fluorescence was subtracted. For weak signals, *e.g.* in the exposure conditions of X = 2 and X = 20, the background correction was accomplished on KaleidaGraph software (Synergy) using a weighted curve fit function. In this case, from the original signal, the weighted fit function was subtracted.

*Confocal microscopy.* Confocal microscopy images were acquired using a spinning disk confocal inverted microscope (Olympus IX81) with a 60× oil-immersion objective (N.A. 1.42) provided with an EMCCD camera (Andor iXon 897), a NanoScanZ z-axis nano-positioning stage (Prior Scientific), and iQ3 software (Andor). Nanoparticles were excited with a 561-nm wavelength laser line, PKH67-labeled Curosurf® was excited at 488 nm, and DAPI at 405 nm. Exposure time, gain and laser intensity were identical in all acquisitions. Image stacks were collected with z-step of 0.25 μm. Image processing was performed with ImageJ.

## 3. Results
### 3.1 Silica nanoparticle-Curosurf® dispersions

The silica nanoparticles were thoroughly characterized [39], and a summary of their physicochemical properties is provided in Table 2. The nanoparticles have a hydrodynamic diameter equal to 60 nm at pH < 6.0, which increases to 1280 nm in PBS, and to 1200 – 1300 nm in serum free culture media. Of note, aggregation in physiological media occurs rapidly (< 1 s) upon dilution in media. The reason is that the electrostatic interactions between nanoparticles are screened following variation in pH and ionic strength. TEM images of the nanoparticles showing a diameter equal to 41.2 nm are provided in **S5**. DLS analysis of nanoparticle size in DMEM and RPMI is provided in **S6**.

**Table 2: Summary of the physicochemical characteristics of nanoparticles.**

| Nanoparticle | Silica, $SiO_2$ |
|---|---|
| [a]Diameter, $D$ | 41.2 nm |
| [b]Diameter, $D_h$ (DI-water, pH < 6.0) | 60.0 nm |
| [b]Diameter, $D_h$ in serum-free DMEM | 1300 nm |
| [b]Diameter, $D_h$ in serum-free RPMI | 1200 nm |
| Surface functional group | Amine |
| Surface charge density | +0.62 $e$ nm$^{-2}$ |
| Zeta potential (DI-water, pH < 6.0) | 50 mV |

[a] TEM; [b] DLS

Considering that native surfactant is not readily available, Curosurf®, which is a clinical surfactant substitute used for treatment of RDS, is a valuable choice for *in vitro* investigations. The chemical composition of Curosurf® and that of native pulmonary surfactant is provided in **S2**. The two surfactant fluids are somewhat similar with respect to lipid composition, but differ in their protein compositions. One example is the absence of protein SP-A in Curosurf® which is associated with tubular myelin formation in native surfactant [18]. Cholesterol which is responsible for less-ordered lipid structures in native surfactant is also absent in Curosurf® [18]. Despite these differences, using cryo-TEM, we show that the lipid structures such as uni- or,

multi- lamellar/vesicular structures with sizes ranging from a few tens of nanometers to about 2 μm are present in Curosurf® [26,39]. These images are presented in **S7**.

The silica nanoparticle-Curosurf® dispersion at $X = 80$ was investigated using TEM. This mixing ratio is smaller than the estimated mixing ratio in occupational exposure which is of the order of $X = 10^4$ [29]. Despite such difference, a lower mixing ratio was used to provide sufficient nanoparticle density for TEM observations. Several examples of 70-nm sections of microscopic samples are shown in Fig. 2. When mixed with nanoparticles, Curosurf® is observed to retain its vesicular structure, *e.g.* the images in Fig. 2 show the presence of uni- and multi-lamellar/vesicular structures, similar to cryo-TEM images of pristine Curosurf® (**S7**). Within the multivesicular structure shown in Fig. 2a, individual vesicles, multi-vesicular vesicles as well as lamellar structures are observed. These structures are distributed in the area confined by the outer membrane, which appears to be disintegrated in the upper part of the image marked with an arrow. The observed fluctuations in the contour of all membranes is associated with surface tension or elastic properties of the membrane. Generally, these fluctuations are expected to occur when the liquid volume inside a membrane is reduced resulting in a reduction in membrane pressure and contour fluctuation [44].

Figure 2b shows a multilamellar vesicle with a size of about 2.2 μm. The overall structure is again well preserved after treatment, and one observes densely packed lamella with gap distances between 20 – 100 nm. This densely packed onion-like structure is reminiscent of multilamellar vesicles found in lipids as well as in surfactant phases [18]. An enlarged section of an inner membranous structure is shown in Fig. 2c in which membrane thickness equal to 5 nm and interlamellar spacing in a range of a few tens of nm are clear. It is noted that the multivesicular/lamellar structures of Fig. 2a and Fig. 2b are devoid of nanoparticles. This is due to a low concentration of nanoparticles at a mixing ratio of $X = 80$.

The multivesicular structure shown in Fig. 2d contains several nanoparticles dispersed throughout its volume with some nanoparticles visible inside the inner membranes giving in total 14 nanoparticles in this 70-nm section. A close inspection of internal nanoparticles reveals that no SLB is formed around them. An enlarged view of selected nanoparticles is shown in Fig. 2e. In this example, it is observed that the nanoparticles come into contact and adhere to the lipid membranes as marked with an arrow on this image. This is resulting from electrostatic interactions between positively charged silica and vesicle membranes: a surface charge density of +0.62 $e$ nm$^{-2}$ drives the electrostatic interactions between the nanoparticles and the negative surface potential of surfactant vesicles, about -54 mV [26]. However in larger interlamellar gaps, nanoparticles are found in proximity but not adherent to the lipids. In Fig. 2f, more examples of nanoparticle-lipid adhesion contacts are marked with arrows.

In general, we find at a mixing ratio of $X = 80$ that nanoparticles interact with surfactant lipids resulting in penetration inside the vesicles or adhesion to membranes. However, nanoparticles are also found in proximity to the membranes but not adhering to them. In accordance with DLS data, by increasing nanoparticle concentration for example at a mixing ratio of $X = 2$, the number of interactions is enhanced resulting in large nanoparticle-surfactant lipid aggregates. It is expected that these interactions can affect the contact of nanoparticles with the alveolar cells. These effects are investigated in below.

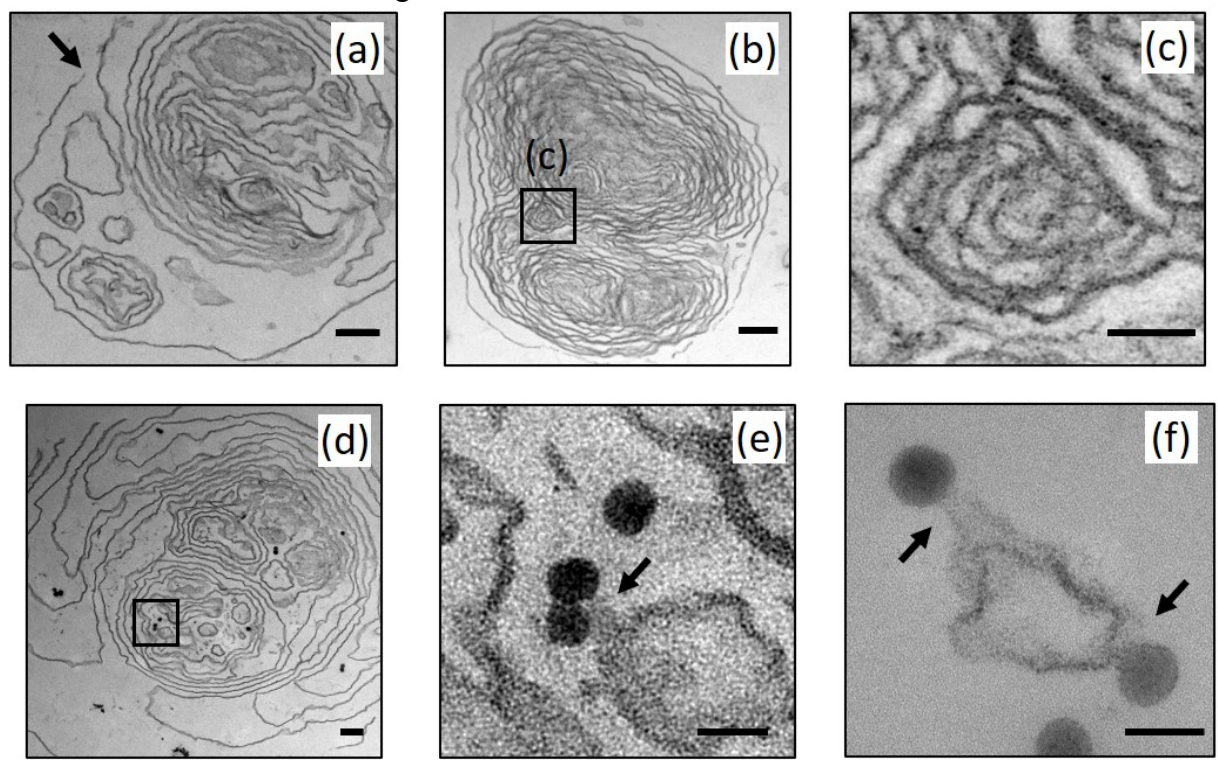

*Figure 2:* Transmission electron microscopy (TEM) images of positively charged silica nanoparticle-Curosurf® dispersion at a mixing ratio equal to $X = 80$. (a) A multivesicular structure containing single vesicles, multivesicular vesicles, and lamellar structures. Arrow points at a disintegration site. (b) A multilamellar vesicle with a size of about 2.2 µm and interlamellar spacing 20 – 100 nm. Boxed area is enlarged in c. (c) An inner structure inside a multilamellar vesicle showing lipid membranes with thickness of 5 nm and interlamellar spacing of a few tens of nm. (d) A multivesicular structure enclosing several nanoparticles in singlets, doublets and triplets. Boxed area shows a single particle and a doublet in the vicinity of lipid membranes. The boxed area is enlarged in e. A total of 18 nanoparticles are found in this 70-nm section. (e) The enlarged boxed area in d. The doublet is in contact with the neighboring lipid membrane as marked with an arrow. (f) Nanoparticles forming adhesion contacts with a lipid membrane are marked with arrows. Scale bare 200 nm (a, b and d), 50 nm (c, e, f).

### 3.2 Integrity of cell barrier

A characteristic feature of epithelial cells is the formation of tight junctions between adjacent cells which prevent diffusion of material across the epithelium. Paracellular permeability is controlled *via* expression, localization and interaction of proteins that form tight junctions such as claudins, occludin and ZO-1 [45]. Using immunostaining, we visualized ZO-1, which anchors the tight junction to actin cytoskeleton, and claudin-4, one of the main sealing proteins that closes the paracellular junction in distal lung epithelium. Examples of the immunostaining images are shown in Fig. 3. From this figure, it was evident that the localization of these key proteins is closely related, but that specific junctional localization is much less consistent in A549 than in NCI-H441. In particular, the distribution of ZO-1 and claudin-4 in A549, respectively in Fig. 3a and 3c, is patchy and dispersed. However, the distribution of these proteins in NCI-H441, respectively shown in Fig. 3b and 3d, is present all around the cells.

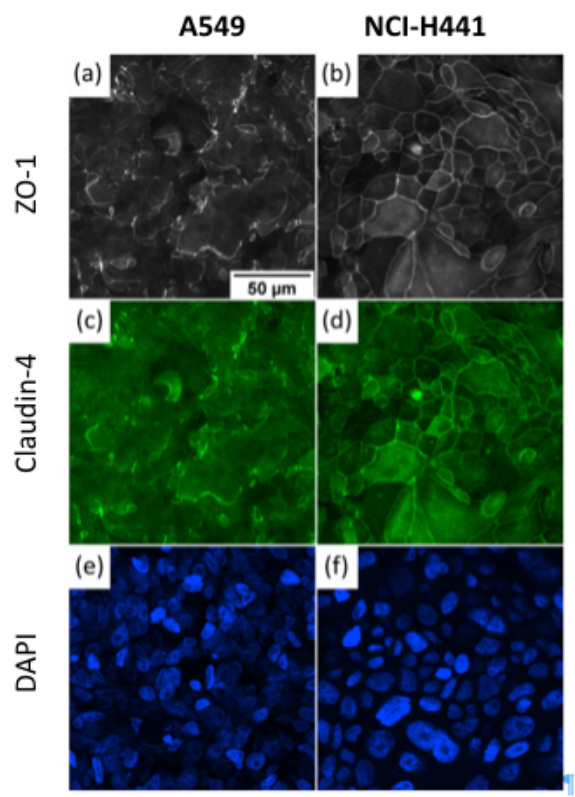

*Figure 3: Distribution of tight junction proteins in A549 and NCI-H441. A549 and NCI-H441 (images on the left and right, respectively) were cultured in ALI condition for two weeks. They were then fixed and immunostained to reveal ZO-1 (a and b) and claudin-4 (c and d). Despite maintaining a compact monolayer (DAPI nuclear staining, e and f), A549 exhibits uneven formation of tight junctions where ZO1 and claudin-4 tightly colocalize along the periphery between adjacent cells. NCI-H441 demonstrates much more consistent tight junction formation. Scale bar 50 µm.*

The integrity of the barrier was monitored using TEER. A high resistance is associated with a barrier that is less permeable to small ions in the solution. We started to monitor the resistance three days post cell seeding, when ALI was first established. The TEER values for A549 are shown in Fig. 4. With A549, the transepithelial resistance did not exceed 40 $\Omega$ cm$^2$ over the course of the cell culture. A low resistance indicates a deficit in the formation of functional tight junctions, consistent with the observed non-uniform distribution of the key proteins (Fig. 3). The TEER values of NCI-H441 are also shown in Fig. 4. With this cell line, an initial reduction in the resistance was obtained which is probably associated with changes in the organization of the cells after the initiation of ALI. Around 7 days post cell seeding (4 days post ALI) the resistance starts to increase gradually. The resistance then reaches a plateau at a value of about 200 – 300 Ω cm² around 15 days post cell seeding (12 days post ALI) consistent with formation of a tight barrier.

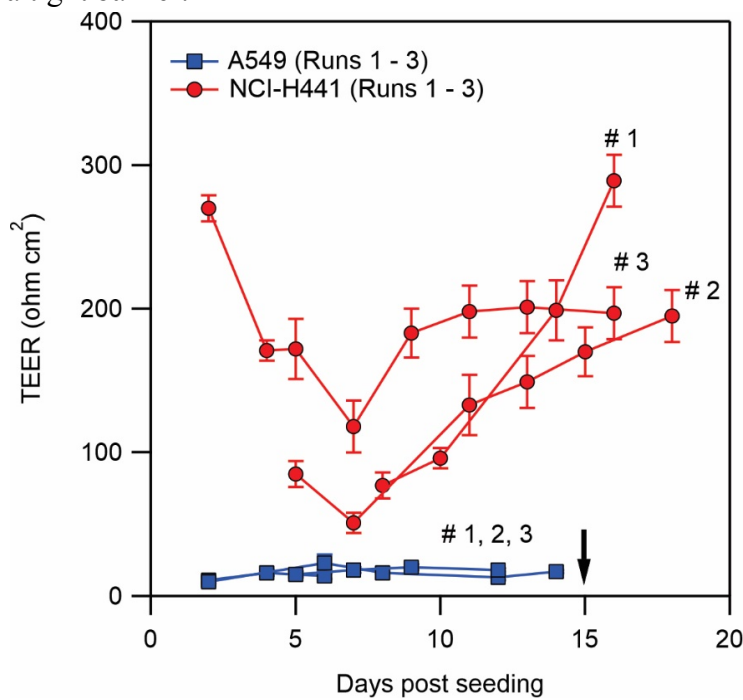

*Figure 4: Transepithelial electrical resistance (TEER) measured for A549 and NCI-H441. ALI was stablished on day 3 post cell seeding when TEER measurements commenced. Nanoparticle exposure was performed after day 15 marked with a black arrow when TEER of NCI-H441 was above 200 Ω cm². Independent experiments were performed on different occasions and in each occasion on at least 3 transwell. Error bar is standard deviation for several transwell in a same experiment.*

### 3.3 Quantification of silica nanoparticle uptake

We investigated the protective role of pulmonary surfactant *in vitro* by exposing A549 and NCI-H441 to dispersions of positively charged silica nanoparticles and Curosurf®. For experiments on coverslips, the cells were exposed to dispersions 2 days after seeding. For experiments on transwell, the cells were exposed generally 15 days post cell seeding (12 days post ALI) when NCI-H441 cells had formed functional epithelial barriers, as evidenced by a TEER ≥ 200 Ω cm². The same time schedule was used for A549 cells cultured on transwell, although the TEER of A549 cells did not exceed 40 Ω cm². The cells were exposed to dispersions for 24 h.

The dispersions included neat silica nanoparticles at $C_{NP} = 100$ µg ml⁻¹, silica-Curosurf® dispersions at X = 2 ($C_{NP} = 100$ µg ml⁻¹, $C_{Curo} = 200$ µg ml⁻¹) and X = 20 ($C_{NP} = 100$ µg ml⁻¹, $C_{Curo} = 2000$ µg ml⁻¹), and neat Curosurf® at $C_{Curo} = 200$ µg ml⁻¹. We note that X = 2 exposure condition corresponds to the peak of light scattering intensity of dispersions signifying a mixing ratio where the interactions result in largest aggregates [26]. The condition X = 20 is closer to the estimated mixing ratio in the lungs [29]. The results of the DLS measurements are presented in **S8**.

Furthermore, the choice of the aforementioned concentrations is driven from the cellular viability assays. We find that in the absence of serum, the onset of cytotoxic effects leading to loss of cellular viability is at $C_{NP} = 100$ µg ml$^{-1}$. No significant loss of viability is observed at lower concentrations with these nanoparticles on exposure tests with A549. In the presence of serum, over a wide range of nanoparticle concentration up to $C_{NP} = 1000$ µg ml$^{-1}$ no loss of viability was observed. No loss of viability was observed if Curosurf® was added to nanoparticle solution at X = 20 even without the presence of the serum. The results of the viability assays are presented in **S9**.

Under the present exposure conditions, the mass of nanoparticles was 200 µg for exposure on coverslip and 50 µg for exposure on transwell. Correspondingly, the surface concentration of nanoparticles was 20.8 µg cm$^{-2}$ on coverslip and 44.6 µg cm$^{-2}$ on transwell. A summary of the exposure conditions is provided in Table 1. A complete overview of the preparation and exposure concentrations is provided in **S3** & **S4**. The investigations were performed using fluorescence microscopy to quantify nanoparticle uptake in each exposure condition. Confocal microcopy was used to localize the silica nanoparticles and Curosurf® lipids at single cell level in addition to semi-quantification of occupied surface area by nanoparticles inside cells.

### 3.3.1 Nanoparticle-cell interactions

After exposure for 24 h, the cells were fixed and observed under fluorescence microscopy for quantification of nanoparticle uptake. Figure 5 shows images of A549 cultured on coverslip (a, b) and on transwell (c, d). In these images, the red fluorescence signal and the phase contrast image are superimposed. On coverslip, the cell layer is more than 90% confluent in selected regions while on transwell the confluency is 100%. The images represent two exposure conditions, namely neat nanoparticle (a, c) and nanoparticle-Curosurf® dispersion at X = 2 (b, d). The images were acquired using a 10× objective with a wide field of view of ~ 0.9 × 0.7 mm$^2$ thus giving access to a large surface area for the quantification. Comparing Fig. 5a (resp. Fig. 5c) to Fig. 5b (resp. Fig. 5d), it is evident that the amount of nanoparticles is reduced when Curosurf® is present. While in the neat exposure condition, a uniform distribution of nanoparticle patches is observed on both coverslip and on transwell, these patches are less frequent in the presence of Curosurf®. To quantify this reduction, we integrated the intensity of the red fluorescence signal for each exposure condition and the resulting values are shown in Fig. 5e and 5f. Thereby, the integrated intensity of samples exposed to the same surface concentration of nanoparticles in the absence of surfactant is found to be much higher than when Curosurf® is present. In particular, a reduction in the integrated intensity of up to two orders of magnitude on glass, *e.g.* ~4000 counts per second (cps) *vs* ~40 cps, and an order of magnitude on transwell, *e.g.* ~13000 cps *vs* ~2000 cps, are observed when Curosurf® is present. Additional increase in the amount of Curosurf® to the exposure condition X = 20 further reduces the amount of nanoparticles retained by A549 cells on transwell, and we conclude that pulmonary surfactant inhibits the uptake of positively charged silica nanoparticles by A549 cells.

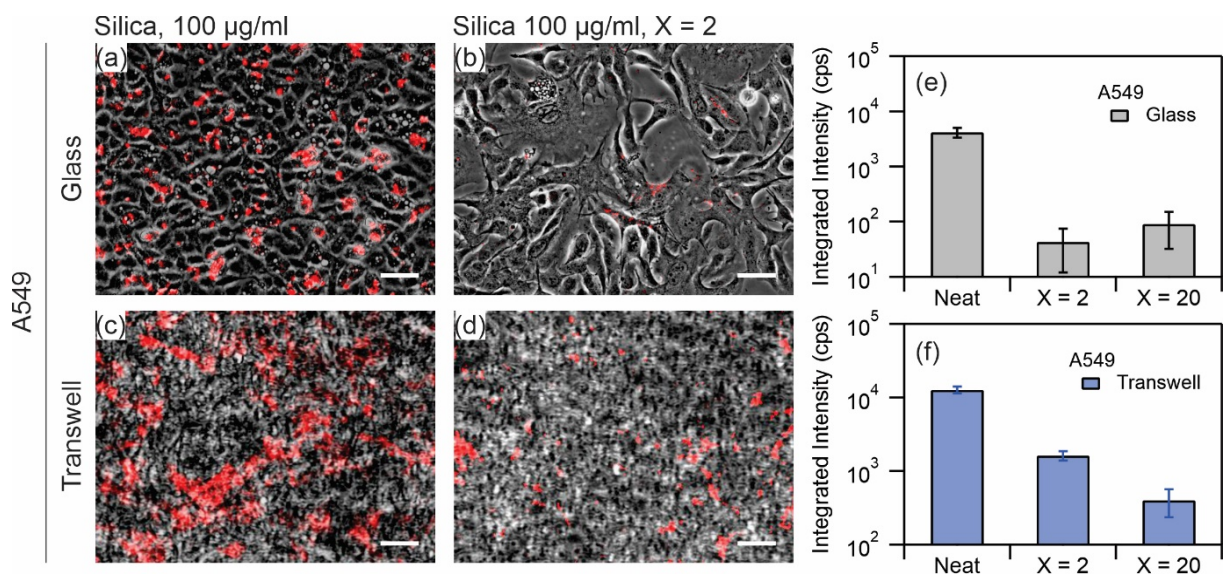

*Figure 5:* (a-d) Superimposed red fluorescence and phase contrast images of A549 exposed to silica nanoparticles at 100 µg ml$^{-1}$ (a, c) and silica nanoparticles at 100 µg ml$^{-1}$ mixed with Curosurf® at 200 µg ml$^{-1}$, X = 2 (b, d), on glass (a, b) and on transwell (c, d). Scale bar 50 µm (a, b) and 25 µm (c, d). (e, f) Quantitative analysis of integrated intensity of the fluorescence signal from the three exposure conditions, namely silica at 100 µg ml$^{-1}$ and silica at 100 µg ml$^{-1}$ mixed with Curosurf® at 200 µg ml$^{-1}$ (X = 2) and 2 mg ml$^{-1}$ (X = 20). Each exposure condition was tested in duplicates.

Figure 6 shows similar images for NCI-H441 cultured on coverslip (a, b) and on transwell (c, d). The images represent the exposure conditions, namely neat nanoparticles (a, c) and silica-Curosurf® dispersion at X = 2 (b, d). As before, the distribution of nanoparticles is visibly different between the two exposure conditions (Fig. 6a *vs* Fig. 6b and Fig, 6c *vs* Fig. 6d), and while clusters of nanoparticle aggregates are observable across the cell layer in exposure to neat nanoparticles, these are less frequent when Curosurf® is present. This observed reduction is consistent with cells cultured on glass and on transwell. The values of the integrated intensity are shown in Fig. 6e for NCI-H441 on glass and Fig. 6f on transwell. A reduction by about two orders of magnitude is observed in the integrated intensity (*e.g.* ~3000 cps *vs* ~50 cps on glass, ~6000 cps *vs* ~90 cps on transwell) between the exposure to neat silica nanoparticles and to silica-Curosurf® dispersion. Such a reduction indicates that the presence of Curosurf® reduces the amount of nanoparticles taken up by NCI-H441, as it did for A549.

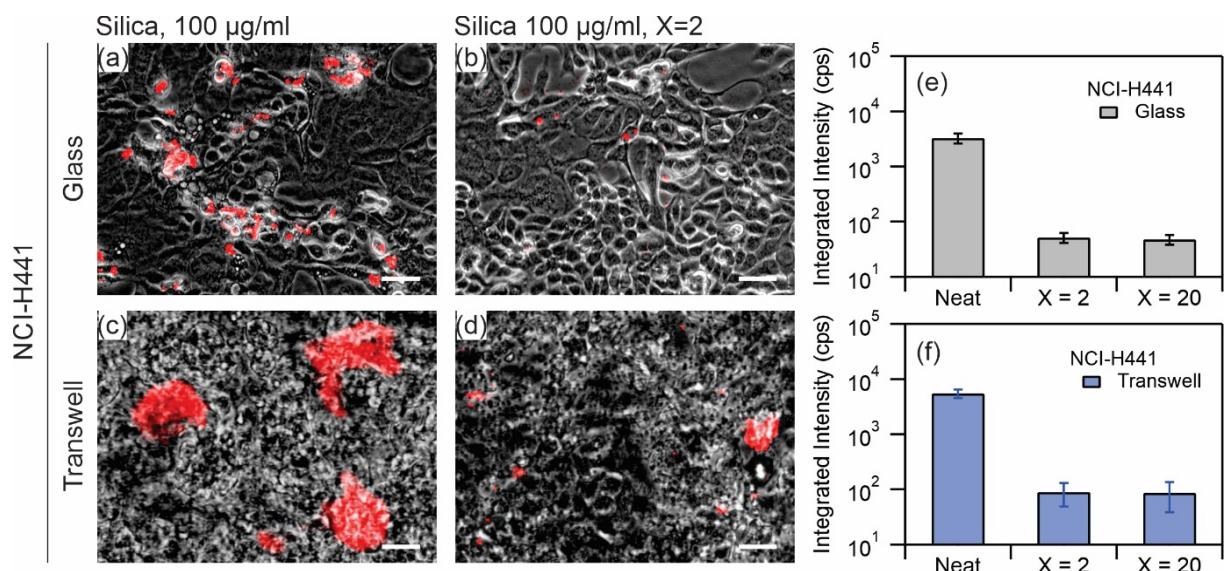

*Figure 6:* (a-d) Superimposed red fluorescence and phase contrast images of NCI-H441 exposed to silica nanoparticles at 100 µg ml$^{-1}$ (a, c) and silica nanoparticles at 100 µg ml$^{-1}$ mixed with Curosurf® at 200 µg ml$^{-1}$, X = 2 (b, d), on glass (a, b) and on transwell (c, d). Scale bar 50 µm (a, b) and 25 µm (c, d). (e, f) Quantitative analysis of integrated intensity of the fluorescence signal from the three exposure conditions, namely silica at 100 µg ml$^{-1}$ and silica at 100 µg ml$^{-1}$ mixed with Curosurf® at 200 µg ml$^{-1}$ (X = 2) and 2 mg ml$^{-1}$ (X = 20). Each exposure condition was tested in duplicates.

The capacity of the two cell lines to adsorb or internalize nanoparticles differs, and whereas A549 exhibited a uniform distribution of nanoparticle fluorescence across the population after 24 h exposure without Curosurf®, NCI-H441 showed a heterogeneous pattern of retention, with isolated cells or patches of cells adsorbing large quantities of nanoparticles while others exhibited little to no retention. This likely arises from underlying differences between the two cell lines in terms of cell uptake processes. Prior research suggests that A549 takes up nanoparticles of this approximate size *via* clathrin- or caveolin-mediated endocytosis, and does not express flotillin proteins [46], whereas NCI-H441 exhibits at least some degree of flotillin-dependent nanoparticle uptake [47]. Despite these differences between cell lines, the integrated intensity of nanoparticle fluorescence is reduced by one to two orders of magnitude when nanoparticles are exposed to cells together with Curosurf® at X = 2, as compared to nanoparticle exposure without surfactant. This reduction again supports a protective role for surfactant against silica nanoparticles. The results of fluorescence microscopy measurements are summarized in Table 3.

### 3.3.2 Localization of silica nanoparticles and Curosurf®
Figure 7 shows confocal microscopy images of A549 cultured on coverslip (Fig. 7a and Fig. 7b) and on transwell (Fig. 7c and Fig. 7d). The representative images are from two exposure conditions, namely neat silica nanoparticles (Fig. 7a and Fig. 7c) and silica nanoparticle-Curosurf® dispersion at X = 2 (Fig. 7b and Fig. 7d). Figure 7a shows a maximum projection of a confocal stack, superimposed over a corresponding phase contrast view of the same region, and a corresponding orthogonal projection to the right of this. The A549 in this image was cultured on a coverslip and exposed to nanoparticles in suspension at a concentration of 100 µg ml$^{-1}$ without surfactant. The phase contrast image shows formation of several vacuoles (1 – 10 µm) in the cells in this exposure condition. We have previously observed a similar morphological response with these cells upon exposure to similar nanoparticles [39]. The vacuoles appear to not contain any nanoparticles or aggregates. This observation suggests that although their formation is part of cellular response to nanoparticle exposure, they do not directly participate in the internalization process. The orthogonal image depicts a section along the direction of triangle and shows that the nanoparticle aggregates are inside the cells and at the level of or below the nuclei.

Figure 7b shows an example of identical cells exposed to silica-Curosurf® dispersion at X = 2. In this case, the intracellular vacuoles are not visible anymore. This observation which shows the effect of pulmonary surfactant on eliminating one possible cytotoxic cellular response is again consistent with the behavior of these cells in exposure to SLB-coated silica nanoparticles [39]. Moreover, we find that the amount of internalized or surface bound nanoparticles is reduced in exposure to X = 2 condition. We also find that the nanoparticles colocalize with surfactant lipids: yellow spots on the superimposed red fluorescence from silica nanoparticles and green fluorescence from Curosurf® lipids confocal stacks. The orthogonal view again shows that the colocalized nanoparticle-lipid aggregates are inside the cells and at the level of the nuclei. Observation of green aggregates reveals that surfactant lipids may also be internalized by the cells or bound to the cell surface. The exposure conditions in Fig. 7c and Fig. 7d are similar to Fig. 7a and Fig. 7b; however, the cells were cultured on transwell in these cases. Again, one observes nanoparticle aggregates when nanoparticles were incubated with the cells (Fig. 7c) and that the number of these aggregates is reduced in the presence of surfactant (Fig. 7d). Similar to Fig. 7b, the observed aggregates in Fig. 7d are found to colocalize with surfactant lipids and are visible as yellow spots.

From these images it is possible to estimate the area fraction of nanoparticles inside the cells in each exposure condition. Figure 7e (resp. 7f) shows that on glass (on transwell) the occupied area by the nanoparticles in the exposure to neat nanoparticles at a concentration of 100 µg ml$^{-1}$ is much higher than when Curosurf® is present by two times more in mass, *i.e.* the exposure condition X = 2. The reduction is persistent on both coverslip and on transwell. Moreover, additional increase in surfactant concentration further decreases the estimated occupied area inside the cells. These observations are consistent with the results of fluorescence microcopy for A549 shown in Fig. 5, and further corroborates the protective role of surfactant fluid against positively charged nanoparticles.

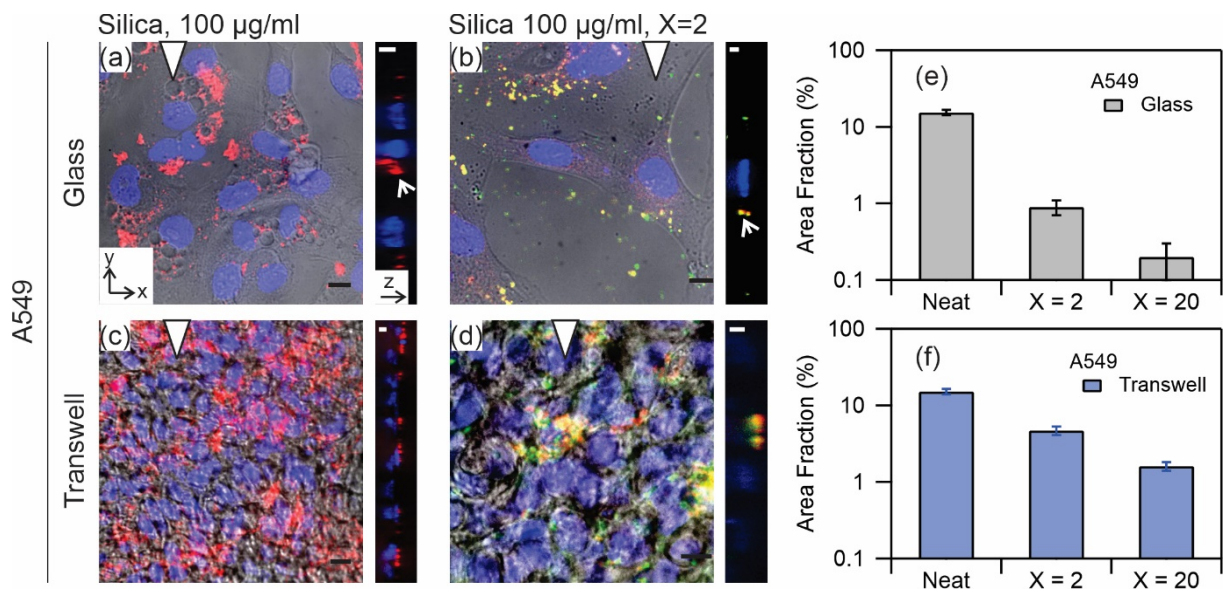

*Figure 7:* Confocal microscopy images of A549 exposure to silica nanoparticles at 100 µg ml$^{-1}$ (a and c) and silica at 100 µg ml$^{-1}$ mixed with Curosurf® at 200 µg ml$^{-1}$, X = 2 (b and d) on glass (a and b) and on transwell (c and d). The x-y plane images are merges of phase contrast and red, green and blue confocal stacks. The colors are respectively from Cy3-tagged nanoparticles, PKH67-tagged Curosurf® and DAPI-tagged nuclei. The presented y-z plane images are merges of red, green and blue confocal stacks, left is bottom and right in top. The triangle positions the y-z plane. Scale bar 10 µm (x-y), and 5 µm (y-z). (e, f) Semi-quantitative analysis of area occupied by nanoparticles inside the cells for three exposure conditions, namely neat silica nanoparticles, and nanoparticles mixed with Curosurf® at X = 2 and X = 20.

Figure 8 (a-d) show similar confocal images for NCI-H441. These cells were again cultured on coverslip (Fig. 8a and Fig. 8b) or on transwell (Fig. 8c and Fig. 8d), and were incubated with neat silica nanoparticles at 100 µg ml$^{-1}$ (Fig. 8a and Fig. 8c) and dispersion of nanoparticles and Curosurf® at X = 2 (Fig. 8b and Fig. 8d). The occupied area fraction is estimated in Fig. 8e and Fig. 8f. Several observations are consistent with A549. The x-y plane image of Fig. 8a shows formation of vacuoles inside the cells in this exposure condition and agrees with our previous observation of this morphological response in A549 [39]. The comparisons between Fig. 8a *vs* Fig. 8b and Fig. 8c *vs* Fig. 8d show that the number of internalized or surface bound nanoparticles is reduced when surfactant is present. The yellow spots in the merged images in Fig. 8b show that the nanoparticles colocalize with pulmonary surfactant lipids. The orthogonal images in Fig. 8a and Fig. 8b show nanoparticle aggregates inside the cell and at the same level as the nuclei, but similar images in Fig. 8c and Fig. 8d show that these aggregates are accumulated in the upper parts of cytosol and above the nuclei. Semi-quantification in terms of area fraction confirms a notable reduction in the occupied area by the nanoparticles. These results further show the protective role of pulmonary surfactant fluid again silica nanoparticles in NCI-H441.

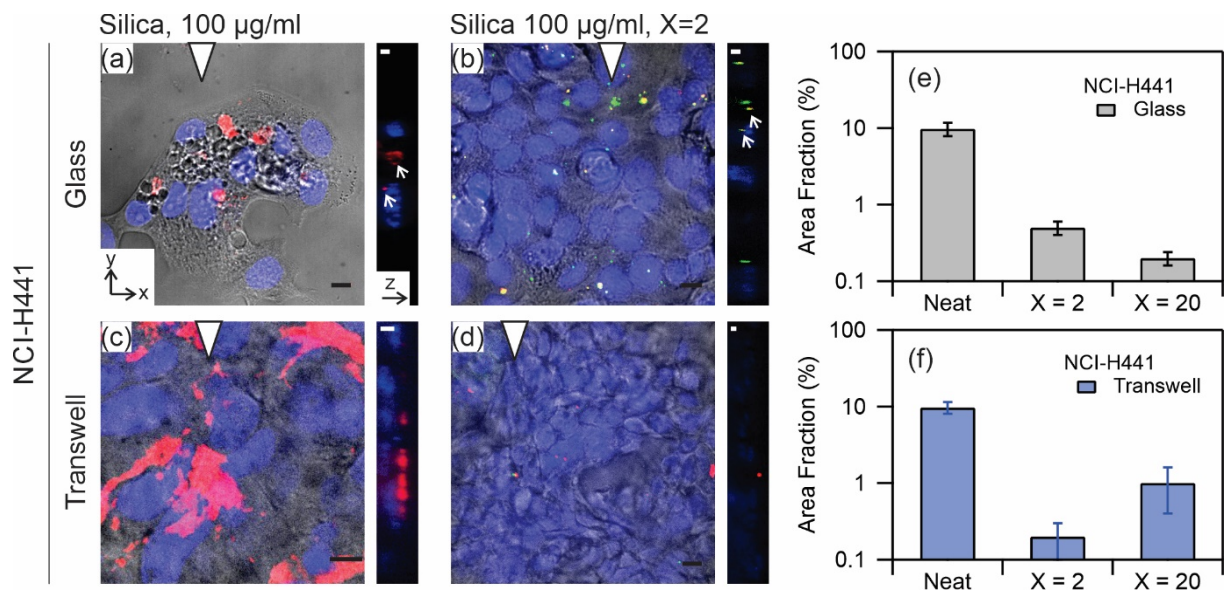

*Figure 8:* Confocal microscopy images of NCI-H441 exposure to silica nanoparticles at 100 µg ml$^{-1}$ (a and c) and silica at 100 µg ml$^{-1}$ mixed with Curosurf® at 200 µg ml$^{-1}$, X = 2 (b and d) on glass (a and b) and on transwell (c and d). The x-y plane images are merges of phase contrast and red, green and blue confocal stacks. The colors are respectively from Cy3-tagged nanoparticles, PKH67-tagged Curosurf® and DAPI-tagged nuclei. The presented y-z plane images are merges of red, green and blue confocal stacks, left is bottom and right in top. The triangle positions the y-z plane. Scale bar 10 µm (x-y), and 5 µm (y-z). (e, f) Semi-quantitative analysis of area occupied by nanoparticles inside the cells for three exposure conditions, namely neat silica nanoparticles, and nanoparticles mixed with Curosurf® at X = 2 and X = 20.

### 4. Discussion

Our group recently showed that nanoparticles with positive surface charges interact strongly with pulmonary surfactant *via* electrostatic forces [26,29,38], and it was speculated that these effects may modulate nanoparticle interactions with cells. In a previous work, we focused our attention on SLB-coated silica nanoparticles [39]. The SLBs were found to significantly reduce cellular internalization of nanoparticles by A549 on glass (**S10**). Without SLBs, the nanoparticles were found to enter the cells *via* a non-endocytosis pathway; evident from the absence of vesicles surrounding the internalized nanoparticles or nanoparticle aggregates. Moreover, at the port of entry to the cells, the membrane was damaged. These effects were found to be absent on exposure to SLB-coated nanoparticles clearly manifesting the effect of surfactant SLB on cell protection [39]. The current study investigates the protective role of surfactant in an attempt to mimic the *in vivo* exposure condition more closely. Thereby, silica nanoparticles and Curosurf® were mixed without sonication. We observed with TEM that the vesicular structures in Curosurf® are maintained are mostly conserved, and that nanoparticles may be internalized inside vesicles and multivesicular structures. Formation of these structures, in addition to adhesion of lipid fragments to nanoparticles, are expected to modulate nanoparticle-cell interactions. A549 and NCI-H441 were then investigated for the exposure tests when cultured on glass and on transwell; the latter substrate was used to stablish ALI and to induce tight junction which is a characteristic of epithelial cells. Indeed, NCI-H441's tight junction proteins were found to be uniformly distributed around the cytoplasm and a high TEER was obtained with this cell line (Fig. 3 and Fig. 4). A549 was devoid of these characteristics. Nevertheless, we presented the results of cellular uptake with both cell lines as A549 is a common model that is largely investigated in the past literature. Our previous work was also performed with A549 which allowed cross-checking the outcomes of the current and the previous study. In the tests, the mass of nanoparticles was 200 µg for exposure on coverslip and 50 µg for exposure on transwell. Correspondingly, the surface concentration of nanoparticles was 20.8 µg cm$^{-2}$ on coverslip and 41.6 µg cm$^{-2}$ on transwell. While the exposed mass is in agreement with the estimated daily lung burden previously discussed, the surface concentration is an overestimation. In fact, recapitulating the large surface area of alveolar region (70 m$^2$)[5] is not possible *in vitro*, and further lowering the concentration of nanoparticles may prone them to be undetectable [29]. Thereby our results present a high nanoparticle burden scenario and the effect of surfactant in alleviating such extreme condition. A summary of the exposure conditions is provided in Table 1. A complete overview of the preparation and exposure concentrations is provided in the **S3** & **S4**.

We note that the concentration of (corona forming) proteins in Curosurf® is minute (**S2**), and thereby while coronas may form on limited number of nanoparticles, the main effect comes from formation of aggregates with the lipids. It is evident from the results of DLS (**S8**) and TEM (Fg. 2) presented earlier; the sizes of the (smaller) aggregates are in the range of a few microns. The presence of Curosurf® may then change the sedimentation profile of the particles. We suspect that the nanoparticle-Curosurf® aggregates have a density that is close to that of the vesicles, whereas the density of the nanoparticle aggregates is closer to the density of the silica. This difference may then lead to dissimilar sedimentation profiles between the exposure conditions with and without the surfactant. This difference may have an impact on the effective nanoparticle dose submitted to the cells.

Quantification of nanoparticle uptake using fluorescence microscopy revealed that pulmonary surfactant inhibits the uptake by A549 and NCI-H441 on coverslip and on transwell (Fig. 5 and Fig. 6). Particularly, a reduction by one to two orders of magnitude is obtained on exposure to nanoparticle-Curosurf® dispersion at X = 2 comparing to neat nanoparticles. A further increase in the concentration of Curosurf® (condition X = 20) did not notably reduce the amount of uptake any further. In accordance with DLS measurements, at X = 2 and higher, all nanoparticles are readily "complexed" with Curosurf® vesicles. A mixing ratio of X = 2 is apparently sufficient to modify the interactions with the cells and further increase in the X value is therefore not significant.

The current work extends our findings to NCI-H441 and to ALI culture on transwell. The results of fluorescence microscopy measurements are summarized in Table 3, and show up to two orders of magnitude reduction in nanoparticle uptake when surfactant is present. Furthermore, confocal microscopy revealed that nanoparticles and surfactant lipids are co-localized inside the cells in agreement with TEM observations of nanoparticles inside the vesicles and adherent to membranes. Quantification of the occupied area by the nanoparticles inside the cells from confocal microscopy is in general agreement with results from fluorescence microscopy. In comparison between the two cell lines, A549 is found to uptake more nanoparticles than NCI-H441. For example, on transwell, only 2% of the particles are uptaken by NCI-H441. This amount is 3 – 11% for A549. Thereby, cells with tight junctions on transwell are "better protected". This finding may be related with the different cellular processes of these cells in interaction with nanoparticles. These differences are indicative of the importance of using relevant alveolar cell models (*e.g.* NCI-H441 *vs* A549) for *in vitro* investigations, although common models such as A549 may provide useful information on large scale effects.

**Table 3: Comparison of cell exposures.**

| Substrate | Glass coverslip | | | | Transwell | | | |
|---|---|---|---|---|---|---|---|---|
| Exposure condition to silica nanoparticles | Neat | X=2 | X=20 | Ratio X=20/Neat | Neat | X=2 | X=20 | Ratio X=20/Neat |
| **NCI-H441** | | | | | | | | |
| Integrated intensity [a] | 3290 | 52 | 48 | 0.015 | 5515 | 90 | 86 | 0.016 |
| Area fraction [b] | 9.8 | 0.5 | 0.2 | 0.020 | 9.7 | 0.2 | 1 | 0.021 |
| **A549** | | | | | | | | |
| Integrated intensity [a] | 4161 | 43 | 91 | 0.022 | 12701 | 1614 | 402 | 0.032 |
| Area fraction [b] | 15.5 | 0.9 | 0.2 | 0.013 | 15.2 | 4.7 | 1.6 | 0.105 |

[a] Fluorescence microscopy, counts of intensity divided by the total time of exposure in seconds (cps)
[b] Confocal microscopy, %

**5. Conclusion**

The primary role of pulmonary surfactant is to lower and regulate the air-liquid surface tension in order to facilitate alveolar tissue dynamics during breathing. Its role as a defense mechanism against inhaled nanoparticles is still unknown. We show that pulmonary surfactant inhibits the uptake of positively charged silica nanoparticles in two models of alveolar epithelial cells, namely A549 and NCI-H441. Reductions by up to two orders of magnitude is obtained from analysis of fluorescence signal. Investigation of confocal stacks shows that the nanoparticles are colocalized with surfactant lipids revealing the direct effect of the latter in cellular uptake in accordance with TEM observations of nanoparticle-surfactant interactions. NCI-H441 is a better model for these investigations which is associated with its formation of tight junctions and a high transepithelial resistance when cultured in ALI on transwell. This cell line was also found to take less nanoparticles when compared to A549 showing that alveolar cells may even be better protected *in vivo*. Our findings show the importance of surfactant inclusion in *in vitro* recreation of alveolar epithelium in inhalation toxicology and therapy investigations.

**Supplementary Information:**

**S1** – Production volume and industrial application for silica nanoparticles (2012)

**S2** – Curosurf® composition and comparison with native surfactant

**S3** – Summary of exposure conditions on glass coverslip

**S4** – Summary of exposure conditions on transwell

**S5** – Cryo-TEM images of silica nanoparticles

**S6** – Dynamic light scattering of silica nanoparticles dispersion in culture media

**S7** – Cryo-TEM images of Curosurf® vesicular structures

**S8** – Dynamic light scattering of silica nanoparticles-Curosurf® dispersions

**S9** – Cellular viability assays

**S10** – TEM of internalized silica nanoparticles without or with Curosurf®

**Acknowledgements:** We thank S. Mornet from Institut de Chimie de la Matière Condenseé de Bordeaux, Université Bordeaux, for synthesis of aminated silica nanoparticle. We thank M. Mokhtari from Hospital Kremlin-Bicêtre, Val-de-Marne, France for the gift of Curosurf®. We thank C. Puisney and F. Mousseau for viability assay data, TEM images of Curosurf®, and DLS data. Agence Nationale de la Recherche (ANR) and Commissariat à l'Investissement d'Avenir (CGI) are acknowledged for financial support of this work through Labex Science and Engineering for Advanced Materials and devices (SEAM) ANR 11 LABX 086, ANR 11 IDEX 05 02. We acknowledge ImagoSeine facility, Jacques Monod Institute, Paris, France, and France BioImaging infrastructure supported by the French National Research Agency (ANR-10-INSB-04, Investments for the future). This research was supported in part by the Agence Nationale de la Recherche under the contract ANR-13-BS08-0015 (PANORAMA), ANR-12-CHEX-0011 (PULMONANO), ANR-15-CE18-0024-01 (ICONS), ANR-17-CE09-0017 (AlveolusMimics) and by Solvay.